\documentclass[aip,reprint]{revtex4-2}

\usepackage{graphicx}
\usepackage{dcolumn}
\usepackage{bm}

\usepackage[utf8]{inputenc}
\usepackage[T1]{fontenc}
\usepackage{mathptmx}
\usepackage{etoolbox}
\usepackage{natmove}
\usepackage[caption=false]{subfig}
\usepackage{chemformula} 
\usepackage{physics}
\usepackage[cal=cm]{mathalfa}
\usepackage[acronym]{glossaries}
\glsdisablehyper
\usepackage[capitalise]{cleveref}

\newacronym{md}{MD}{Molecular Dynamics}
\newacronym{qm}{QM}{Quantum Mechanics}
\newacronym{csdf}{CSDF}{Center of the Spin Density Function}
\newacronym{vebe}{VEBE}{Vertical Electron Binding Energy}
\newacronym{dft}{DFT}{Density Functional Theory}
\newacronym{oml}{OML}{Orbital Motion-Limited}

\newcommand{\cvp}{\varepsilon_{\mathrm{0}}}

\renewcommand{\Phi}{\varPhi}

\begin{document}

\title{Deformation Kinetics of Charged Nanoclusters via Surface Diffusion}

\author{Efstratios M. Kritikos}
\email{emk@caltech.edu}
\affiliation{Department of Applied Physics and Materials Science, California Institute of Technology, Pasadena, CA 91125, United States}
\affiliation{Department of Physics, University of Oxford, Parks Road, Oxford, OX1 3PU, United Kingdom}

\author{William A. Goddard III}
\affiliation{Materials and Process Simulation Center, California Institute of Technology, Pasadena, CA 91125, United States}
\affiliation{Department of Applied Physics and Materials Science, California Institute of Technology, Pasadena, CA 91125, United States}

\author{Paul M. Bellan}
\affiliation{Department of Applied Physics and Materials Science, California Institute of Technology, Pasadena, CA 91125, United States}

\begin{abstract}
Low-temperature charged nanoclusters deform, primarily via surface diffusion, when surface electrostatic stresses and surface tension stresses are not balanced. We derive an analytical model for the kinetics of this process, based on the cluster size, net charge, surface tension, local curvature, temperature, and diffusivity. The model predicts the deformation time when either surface tension or electrostatic stress dominates, leading respectively to the collapse or elongation of the cluster. Comparison with molecular dynamics simulations of ice nanoclusters with solvated electrons shows quantitative agreement with a single fitted parameter. The model can predict the deformation kinetics of charged nanoclusters in atmospheric aerosols and complex plasmas.
\end{abstract}

\maketitle

Charged nanoclusters exist in astrophysical plasmas within interstellar clouds and accretion disks \cite{Shukla2015}. It has been found that such clusters consist mainly of water-based ice and are elongated \cite{Davis1951}. The origin of this elongation remains unknown. We recently proposed\cite{Kritikos2025} that deformation initiates at nanosizes, when surface electrostatic stresses due to electron-electron repulsions overcome surface tension stresses (Rayleigh instability \cite{Rayleigh1882}). Then, gaseous water molecules are attracted to the tips of the elongated clusters, due to the strong electric field gradients there, and thus grow into more elongated shapes. The timescales of this elongation, however, remain unknown.

Ice clusters also exist in mesospheric terrestrial clouds and thunderstorms, have a negative charge, and can be elongated \cite{Hervig2009,Gaskell1977}. Formation of negatively charged nanoclusters has also been observed in the plasmas used during semiconductor fabrication \cite{Selwyn1995} and in tokamak fusion reactors \cite{Beckers2023}. When such clusters settle onto a chip during the fabrication process, they reduce manufacturing yield. Understanding their deformation kinetics and growth will help develop efficient methods to eliminate contamination. Finally, elongation of microbes has been proposed as a new strategy to deform and kill bacteria \cite{Laroussi2003}. Knowing the deformation timescales will shed light on the antibacterial properties of plasmas and how to exploit them. 

The elongation kinetics of charged macroscale droplets due to the aforementioned Rayleigh instability have been extensively investigated in the literature. When strong surface electrostatic stresses are present, spherical liquid droplets deform into ellipsoids \cite{Betelu2006,Giglio2020}, and may form Taylor cones\cite{Taylor1964} under strong electrostatic repulsions. These studies\cite{Betelu2006,Giglio2020} focus on regimes where viscous forces drive the deformation and neglect nanoscale surface effects. However, surface diffusion should dominate over viscous flow or bulk diffusion at low temperatures and at nanosized clusters. The reason is that surface molecules are undercoordinated. For water, this undercoordination leads to an activation energy for surface diffusion that is half that of bulk diffusion \cite{Murphy1953,Kimmel2025}. Nanoclusters have a high surface-to-volume ratio, and thus the higher mobility of surface molecules governs the diffusivity. Additionally, considering an Arrhenius-type diffusion coefficient, the lower activation energy of surface molecules leads to an increased surface  over bulk diffusivity as the temperature decreases. For these reasons, this study focuses on deformation driven by surface diffusion. To our knowledge, an analytical model that predicts the timescale of nanocluster elongation driven by surface diffusion and electrostatic repulsions has not been developed yet.

This work derives such an analytical model and validates the model via \gls{md} simulations of charged ice nanoclusters with solvated electrons. The analysis is focused on ice nanoclusters because of their relevance in atmospheric aerosols and astrophysical plasmas; however, the findings are expected to be applicable to any other type of clusters.

The elongated nanoclusters are approximated as ellipsoids (prolate spheroids), having major and minor radii $a$ and $b$, respectively. This assumption is consistent with \gls{md} simulations of charged nanoclusters deforming at low temperatures \cite{Kritikos2025,Kritikos2026}, and with experimental observations of clusters in laboratory plasmas \cite{Nicolov2024}. The aspect ratio is then $\alpha=a/b$. Also, the radius of a sphere with the same volume as the ellipsoid is $R=a^{1/3} b^{2/3}$. Thus, $a=\alpha^{2/3}R$ and $b=\alpha^{-1/3}R$. The total curvature of the ellipsoidal surface is $K=R_1^{-1}+R_2^{-1}$, where $R_1$ and $R_2$ are the two principal radii. 

The contribution of a stress $\tau$ to the chemical potential $\mu$ is given by $\mu=\tau \Omega$, where $\Omega$ is the molecular volume \cite{Mullins1957}. The stress on a surface due to surface tension $\gamma$ is equal to $\tau_{\gamma}= \gamma K$ \cite{Taylor1964}. Combining these relations gives the Gibbs-Thomson equation, $\mu_{\gamma}=\Omega\gamma K$. The chemical potential decreases when molecules or atoms move from a region of high curvature to a region of low curvature (i.e., flatter region). 

The surface electric field of the cluster also affects the total chemical potential as it imposes an outward-facing stress. To include the electric field contribution to the chemical potential, we consider the electrostatic Maxwell stress. The electrostatic stress is equal to $\tau_\mathcal{E}=\epsilon_0 \mathcal{E}^2/2$, and thus the corresponding chemical potential is $\mu_\mathcal{E}=\epsilon_0 \Omega \mathcal{E}^2/2$.

Considering that the surface tension stress and electrostatic stress have opposite directions, the total chemical potential of the surface molecules is $\mu = \Omega(\tau_\gamma-\tau_\mathcal{E})$ or
\begin{equation}
    \mu = \Omega \left( \gamma K - \frac{1}{2} \epsilon_0 \mathcal{E}^2 \right) .
\end{equation}

The surface tension $\gamma$ differs from its planar value as cluster curvatures approach the nanoscale. Tolman\cite{Tolman1949} introduced a radius-dependent correction for nanosized spherical droplets, while Helfrich\cite{Helfrich1973} later derived a correction for a surface with an arbitrary shape and curvature. The two definitions can be combined, giving
\begin{equation}
    \gamma = \dfrac{\gamma_\infty}{1+\delta K},
    \label{eq:gamma}
\end{equation}
where $\delta$ is the Tolman length, which depends on the substance of the cluster. Several different values have been reported for water, both positive and negative, and ranging between $-2.0$ and $1.5$ {\AA} \cite{Kritikos2025}. No consensus has been reached yet on its magnitude, or even its sign. The analytical model presented here is evaluated with a $\delta=-0.8$ \AA, as computed for water clusters with solvated electrons at 100 K via \gls{qm} computations \cite{Kritikos2025}.

For an ellipsoid, the principal radii are equal to $R_1=\alpha^{-1/3}R^{-2}[\alpha^{4/3}R^2+(\alpha^{-2}-1)z^2]^{3/2}$ and $R_2=\alpha^{-1}[\alpha^{4/3}R^2+(\alpha^{-2}-1)z^2]^{1/2}$ \cite{Kritikos2026}. Therefore, the total curvature is
\begin{equation}
    K=\dfrac{\alpha^{1/3}R^2+\alpha[\alpha^{4/3}R^2+(\alpha^{-2}-1)z^2]}{[\alpha^{4/3}R^2+(\alpha^{-2}-1)z^2]^{3/2}}.
    \label{eq:tot_curvature}
\end{equation}

Moreover, assuming that the cluster surface is an equipotential, the electric field normal to the surface of an ellipsoid is \cite{Bellan2020}
\begin{equation}
  \mathcal{E} = \frac{Q}{4\pi\cvp \alpha^{1/3} R \sqrt{R^2-z^2\alpha^{-4/3}\left(1-\alpha^{-2}\right)}}.
\end{equation}
Charged ice nanoclusters have solvated electrons and thus their surface may differ from a conductive distribution. However, we previously showed that this assumption agrees well for the deformation of ice clusters when solvated electrons are distributed near the surface\cite{Kritikos2025}.

We assume that deformation follows the Nernst-Einstein relation. The mobility of particles on the surface of the cluster due to an external force $F_s$ is $\sigma_s=V_s/F_s$, where $V_s$ is the drift velocity. The Nernst-Einstein relation links the surface diffusivity $D_s$ with the mobility of the surface atoms $\sigma_s$ as $D_s = \sigma_s \kappa T$. The driving thermodynamic force for surface diffusion is the chemical potential gradient, i.e., $F_s = -\nabla_s \mu$, where $\nabla_s$ is the surface gradient. Combining the above, we get the average drift velocity of surface molecules
\begin{equation}
V_s = -D_s \nabla_s \mu/ \kappa T.
\end{equation}
The surface current density of molecules' migration is then $J_s= \zeta V_s$, where $\zeta$ is the number of surface molecules per unit area. Next, we find the speed $v_s$ at which a surface element moves along its normal \cite{Mullins1957}. First, we consider a surface element with area $A$ and height $h$. The number of molecules $N$ in that surface element is $N=A h/\Omega$. Thus $h=\zeta\Omega$ since $\zeta=N/A$. The velocity normal to the top surface of the surface element is defined as $v_s=\partial h/\partial t =\Omega \partial \zeta/\partial t$. Mass conservation gives $\partial \zeta/\partial t=-\nabla_s\cdot(\zeta \bm{V}_s)$, and thus $v_s=-\Omega\nabla_s\cdot(\zeta \bm{V}_s)$. Therefore,
\begin{equation}
  v_s = \frac{\Omega^2 \zeta D_s}{\kappa T} \nabla_s^2 \left( \gamma K - \frac{1}{2} \epsilon_0 \mathcal{E}^2 \right),
  \label{eq:normal_surf_vel}
\end{equation}
where a spatially constant $\zeta$ is assumed.

An arbitrary shape would require computation of the exact surface Laplacian. For our analysis, where the ellipsoid is a prolate spheroid, the path that the diffusing surface atoms follow must be the arc from the tip of the ellipsoid, with the highest $\mu$ value, to the equator, with the lowest $\mu$ value, assuming that the ellipsoid deforms uniformly. Any chemical potential differences along the perpendicular direction to the arc $s$ are neglected. The equation for the surface Laplacian is computed from the Laplace–Beltrami operator, which for a prolate spheroid in cylindrical coordinates is \cite{Zhang2023}
\begin{equation}
\nabla_s^2 \mu = \dfrac{\partial^2 \mu}{\partial s^2} + \frac{1}{r} \dfrac{d
r}{d s } \dfrac{\partial \mu}{\partial s } + \frac{1}{r^2} \dfrac{\partial^2
\mu}{\partial \phi^2}
\label{eq:LB_operator}
\end{equation}
where $s$ is the arc length of the path that particles follow when diffusing. As mentioned before, the $\mu$ dependence on $\phi$ is neglected. Each term in \cref{eq:LB_operator} is expressed with respect to $z$, using $\dd s/\dd z = \sqrt{1 + (\dd r/\dd z)^2}$ (cf. Section S1.1 for more details).

Next, we compute the time it takes for the nanocluster to elongate. The velocity of a surface element at position $z$ of the ellipsoid's surface is $\dd z/\dd t = v_s n_z$, where $n_z=z \alpha^{-2} [R^2\alpha^{-2/3}+z^2(\alpha^{-4}-\alpha^{-2})]^{-1/2}$ (cf. Section S1.2) is the $z$-component of the normal unit vector to the surface. We assume that the nanocluster remains a perfect ellipsoid during deformation; thus, it is sufficient to compute the deformation time at a single point. We consider the tip of the ellipsoid, where $z \rightarrow a(t)=\alpha^{2/3}R$, and the velocity $v_s$ points in the $z$ direction when $\alpha$ is changing over time. At this location, $v_s=(2/3)\alpha^{-1/3}R\dd \alpha/\dd t$ so $\dd t/\dd \alpha=2\alpha^{-1/3} R/(3 v_s)$. The time to evolve from $\alpha_1$ to $\alpha_2$ is then
\begin{equation}
  t_d = \int_{\alpha_1}^{\alpha_2} \frac{2 R \alpha^{- 1 / 3} \kappa T}{3 \Omega^2 \zeta D_s \nabla_s^2 \left( \gamma K - \frac{1}{2} \epsilon_0  \mathcal{E}^2 \right) |_{z \rightarrow \alpha^{2 / 3} R}} \dd \alpha.
  \label{eq:deform_time}
\end{equation}
The molecular volume $\Omega$ can be computed directly from the volume of the cluster as $\Omega = 4 \pi R^3/(3 N_m)$. To find the number of molecules per unit surface, we assume that each particle is spherical (i.e., $\Omega = 4 \pi r_p^3/3$), thus the surface area of each particle's projection (circle) is $A_p = \pi r_p^2$. The number of surface atoms per unit area is then $\zeta = 1/A_p=\pi^{-1} [3 \Omega/(4 \pi)]^{- 2/3}$. The actual number will be slightly less because this $\zeta$ formulation assumes perfect packing of circles over the surface area. 

\begin{figure}[!tb]
    \centering
    \includegraphics[width=\linewidth]{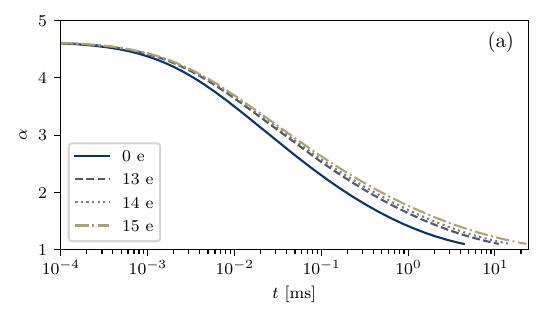}\\[-7pt]
    \includegraphics[width=\linewidth]{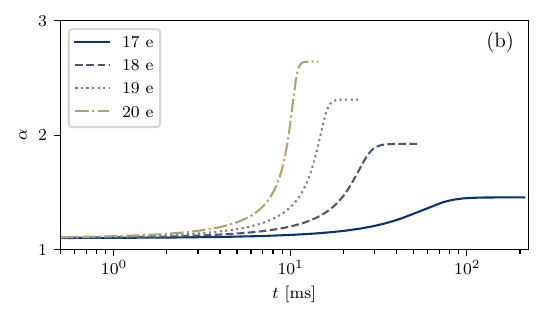}
    \caption{Deformation time computed via \cref{eq:deform_time} for an $R=2.4$ nm cluster at different $Q$ values: (a) from elongated to spherical and (b) from spherical to elongated. The conditions are $T=150$ K, $\gamma_{\infty}=76$ mN/m, $\delta=-0.8$ \AA, $\rho=1$ g/cm$^3$, $D_s=10^{-15}$ m$^2$/s.}
    \label{fig:vel_map_times}
\end{figure}

Based on the above analysis, we can compute the deformation time while the cluster collapses due to a lack of sufficient electrons or elongates when electrostatic stresses are strong enough. Both collapse and elongation are governed by the same \cref{eq:deform_time} and the same surface mobility, and only differ in the sign of the driving force. A parametric analysis of \cref{eq:deform_time} is given in Section S1.3. Timescales are shown in \cref{fig:vel_map_times} (cf. Fig. S1 for the tip velocity). First, an ellipsoidal nanocluster deforming (collapsing) to a sphere due to a lack of electrons (cf. \cref{fig:vel_map_times}a) starts at an accelerated rate and slows down as a more spherical shape is achieved. This is expected because the driving force, which is the surface Laplacian of the chemical potential, decreases with decreasing $\alpha$. When electrons are present on the surface, they slow down the deformation. For the example case in \cref{fig:vel_map_times}, the deformation time increases by 5 times when 15 electrons are added. 

When electrostatic stresses overcome the surface tension (cf. \cref{fig:vel_map_times}b), the deformation timescale trend differs. Initially, $\alpha$ increases slowly and then accelerates almost exponentially. The whole deformation trend resembles a singularity or power-law divergence curve. Notice that at high aspect ratios, the radius of curvature becomes comparable to the $\delta$ value, leading to a strong surface tension stress. Therefore, further elongation takes a significantly longer time. This trend does not exist for positive $\delta$ values, in which case $\gamma$ decreases with increasing curvature, and thus elongation is aided.

We also find that, for strongly negative $\delta$ and large $\alpha$, the tip and nearby surface points may move in opposite directions, suggesting that surface diffusion drives the cluster toward nonellipsoidal shapes. The surface Laplacian is the driving force for deformation via surface diffusion and is shown in Fig. S5. As $\alpha$ increases, the surface Laplacian at the tip of the ellipsoid decreases until eventually it reaches zero. At this point, the tip does not want to move further because the surface tension stresses are too strong. Further increase in $\alpha$ brings the surface Laplacian to negative values, suggesting that the tip wants to move towards the center of the ellipsoid. However, points near the tip have a positive surface Laplacian value, and thus will move outwards even though the tip wants to move inwards. Therefore, the shape will stop being ellipsoidal, likely becoming cylindrical or developing more complex structures. Materials with a slightly negative or positive $\delta$ will not experience this trend.

The deformation timescale, presented here, can explain cluster (grain) evolution in astrophysical plasmas. Clusters in such plasmas exist at low temperatures around 100 K \cite{Nicolov2024}. Electrons and ions will collide with the clusters with velocities near the plasma temperature and will heat them. The deformation time of \cref{eq:deform_time} is affected by the temperature of the cluster via the ratio $T/D_s$. Assuming that the diffusivity has an Arrhenius dependence, the ratio becomes $(T/D_{s,0})\exp(E_{s,a}/(\mathcal{R}T))$, where $D_{s,0}$ is the pre-exponential factor, $E_{s,a}$ is the activation energy, and $\mathcal{R}$ is the ideal gas constant. Low-temperature ice has $E_{s,a}\approx22$ kJ/mol (equivalent to the energy of a hydrogen bond) \cite{Murphy1953}. Thus, an increase from 100 to 120 K would give a decrease in the $T/D_s$ ratio by a factor of $\sim70$, significantly decreasing $t_d$. Ion bombardment in plasmas will also speed up surface diffusion \cite{Bellan2020}. Therefore, deformation is easy for nanoscale grains.

The clusters will continue to grow as water molecules attach at the tip of the ellipsoid, until eventually electrostatic stresses are too weak to affect the shape\cite{Kritikos2025}. However, the cluster temperature is now not significantly affected by high-energy collisions. Specifically, the temperature increase due to a collision energy $Q_c$ is proportional to $Q_c/R^3$. As an example, for an argon plasma with neutral density $n_n=10^{21}$ m$^{-3}$, a $10^{-6}$ fractional ionization, $T_e=2$ eV, and $T_i=30$ meV, a 2.5 nm cluster will experience an increase of $\sim20$ K from an ion collision, whereas the temperature of a 5 nm cluster will increase only by 2 K. Additionally, according to the \gls{oml} theory, the successive collision time between electrons or ions and the cluster is inversely proportional to its surface area $R^2$ \cite{Shukla2015}. 

Overall, small clusters in astrophysical plasmas are expected to deform quickly due to frequent temperature increases by collisions, continue to grow into more elongated shapes as water molecules attach at the tips, until electrostatic stresses are too weak to affect the shape. However, the cluster temperature is no longer significantly affected by high-energy collisions, and thus clusters maintain their elongated shape due to slow surface diffusion.

To evaluate the analytical model, we perform \gls{md} simulations of an $R=2.5$ nm cluster with 2181 water molecules and explicit solvated electrons. Water interactions are modelled with the SPC/E potential, while electron-water interactions are modelled with the potential presented in Ref. \citenum{Kritikos2025}. For the simulation results presented below, the temperature of the nanocluster was 200 K. At this temperature, the surface diffusion of SPC/E water is $\sim7$ times larger than bulk diffusion\cite{Brodskaya2007}, allowing us to directly assess our model via the \gls{md} simulations. The \gls{md} methodology is detailed in Section S2.1. The elongated clusters are prepared through successive heating-cooling cycles\cite{Kritikos2026}. This strategy ensures that the temperature is low enough for the clusters to deform uniformly towards ellipsoidal shapes, yet high enough for the deformation to happen at timescales that are accessible by \gls{md} simulations. 

Although this approach achieves the desired shapes, it cannot be used to extract the elongation kinetics, because the deformation does not occur at a single temperature. Simulating at a constant temperature of 220 K does achieve elongation on MD accessible timescales\cite{Kritikos2025}, but this temperature is above the melting point of SPC/E ($\approx 215$~K)\cite{Vega2005}, where bulk diffusion becomes comparable to surface diffusion\cite{Brodskaya2007} and viscous flow contributes to the deformation. Therefore, we investigate here the collapse of a cluster to assess the analytical model. Specifically, we start from an elongated cluster, remove electrons, and follow its relaxation toward a sphere at constant temperature. As mentioned above, collapse and elongation are governed by the same equation, only differing in the sign of the driving force. Thus, the collapse simulations test both the transport kinetics and the charge dependence of the driving force. Each simulation starts from the elongated cluster with 16 electrons, equilibrated at 200~K. At $t=0$ ns, we randomly select and remove 1, 2, 3, or all 16 electrons. The cluster then deforms under the  canonical (NVT) ensemble at 200~K for 20~ns. 

Each charge state is simulated three times with different initial velocities and electron selections. The aspect ratios are computed directly from the particle trajectories and are averaged over the three simulations. Specifically, we construct the surface mesh of the cluster, compute the cluster volume $V$, and find $\alpha$ from the relations $R^3=3 V/(4\pi)$ and $b^2=R^3/a$ (cf. Section S2.2 for more details).

The analytical model is evaluated using \cref{eq:deform_time} with parameters computed for the SPC/E water model. Wang et al. \cite{Wang2019} computed $\gamma_{\infty}$ of the SPC/E model from \gls{md} simulations. Fitting their data at 200 K gives $\gamma_{\infty}=76.94$ mN/m. For the diffusivity, we use $D_s=2\times10^{-10}$ m$^2$/s as measured by Brodskaya~\cite{Brodskaya2007} for nanoclusters at 200 K with the SPC/E model. A $\delta$ value close to the supercooled regime has not been reported for the SPC/E model. Recently, Chen et al.\cite{Chen2025} found $\delta \approx -0.48$ {\AA} at 300~K with measurements ranging from $-0.35$ to $-0.64$ {\AA}. It has also been reported that the magnitude of $\delta$ for SPC/E decreases with decreasing temperature \cite{Sedlmeier2012}. Since a low-temperature $\delta$ value is not available, we treat it as a fitting parameter.

\begin{figure}
    \centering
    \includegraphics[width=\linewidth]{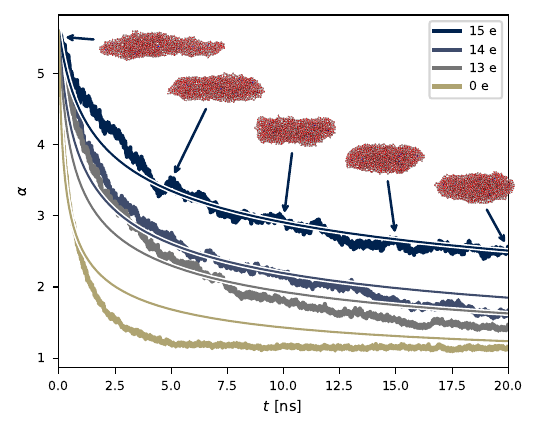}
    \caption{Deformation time of an elongated nanocluster with $R=2.5$ nm, $Q=-16$ e, and initial $\alpha\approx 5.6$. The smooth lines correspond to the theoretical model predictions of \cref{eq:deform_time}, whereas the jagged lines correspond to MD simulations.}
    \label{fig:md_time}
\end{figure}

\Cref{fig:md_time} shows the time evolution of the aspect ratio of an elongated nanocluster after a different number of electrons is removed. Adjusting $\delta$ by hand gives $\delta=-0.05$ {\AA}. With this single value, the model is in good quantitative agreement with the MD results for every number of removed electrons. Specifically, the time it takes for $\alpha$ to drop to half its initial value is within a 1 ns difference between the simulations and the model. Moreover, at the end of the 20 ns simulation time, the model and MD values of $\alpha$ differ by less than 0.2. This agreement is important because changing $\delta$ from $-0.05$ to $-0.1$~{\AA} mainly shifts the curves along the time axis and only slightly changes their shape (cf. Fig. S4). Moreover, the curve with all electrons removed (i.e., the 0 e case) is not considerably affected by $\delta$, so its agreement with the MD results relies on the other parameters that were independently reported in the literature (i.e., excluding $\delta$).

The model correctly captures the deformation time curve and the temporal impact of removing more electrons from the nanocluster. When few electrons are removed, the electrostatic stresses are significant enough to slow down the deformation due to the surface tension stresses. This slowing is captured well with the same $\delta$ for all charge conditions, since the charge only affects via $\mathcal{E}$. When all electrons are removed, the nanocluster becomes spherical slightly faster in MD compared to the analytical model prediction. This faster deformation is attributed to the removal of the electrons, which leaves 16 empty cavities, and thus allows for bulk diffusion to contribute to the deformation. Note also that these simulations predict a lower $|\delta|$ value compared to the ones performed at room temperature by Chen et al.\cite{Chen2025}. The small magnitude of the fitted value is attributed to two reasons: First, $|\delta|$ decreases significantly as the temperature drops~\cite{Sedlmeier2012}. Second, because $\delta$ was originally derived for equilibrium droplets \cite{Tolman1949}, it may vary during the non-equilibrium deformation process. In the latter case, the Tolman length may be a kinetics-dependent property, which should be investigated in future work.

Overall, the analytical model agrees with the MD simulations and thus can describe the deformation time of elongated nanoclusters. Using this model, the nanocluster growth and deformation in complex plasmas and atmospheric aerosols can be predicted.


\begin{acknowledgements}
PMB thanks the US National Science Foundation (NSF) for funding (grant number 2308558). WAG thanks the Liquid Sunlight Alliance (LiSA) for support, which is supported by the US Department of Energy, Office of Science, Office of Basic Energy Sciences, Fuels from Sunlight Hub under Award Number DE-SC0021266. The computations presented here were conducted in the Resnick High Performance Computing Center, a facility supported by Resnick Sustainability Institute at the California Institute of Technology.
\end{acknowledgements}


\bibliography{references}

\end{document}


\title{Deformation Kinetics of Charged Nanoclusters via Surface Diffusion -- Supporting Information}

\author{Efstratios M. Kritikos}
\email{emk@caltech.edu}
\affiliation{Department of Applied Physics and Materials Science, California Institute of Technology, Pasadena, CA 91125, United States}
\affiliation{Department of Physics, University of Oxford, Parks Road, Oxford, OX1 3PU, United Kingdom}

\author{William A. Goddard III}
\affiliation{Materials and Process Simulation Center, California Institute of Technology, Pasadena, CA 91125, United States}
\affiliation{Department of Applied Physics and Materials Science, California Institute of Technology, Pasadena, CA 91125, United States}

\author{Paul M. Bellan}
\affiliation{Department of Applied Physics and Materials Science, California Institute of Technology, Pasadena, CA 91125, United States}

\begin{abstract}
This document reports additional derivations and results in support of the discussion presented in the main paper. The structure of the Supporting Information is as follows:
\tableofcontents
\end{abstract}

\maketitle

\pagebreak

\section{Elongation kinetics}
\label{sisec:kinetics}

\subsection{Surface Laplacian}
\label{subsec:surf_lapl}

The terms in the Laplace-Beltrami operator in the manuscript are expressed with respect to $z$, so $\dd r/ds=(\dd r/\dd z)/(\dd s/\dd z)$. Using the equation for the ellipse,
\begin{equation}
r=\dfrac{R}{\alpha^{1/3}}\sqrt{1-\dfrac{z^2}{\alpha^{4/3}R^{2}}},
\label{sieq:surf_lapl_temp1}
\end{equation}
the $dr/dz$ term can then be determined. Also, for our cylindrical coordinates, $\dd s/\dd z = \sqrt{1 + (\dd r/\dd z)^2}$. Therefore,
\begin{equation}
\frac{\partial \mu}{\partial s} = \left[ 1 + \frac{z^2}{\alpha^{10 / 3} {R^2}} \left( 1 - \frac{z^2}{\alpha^{4 / 3} R^2} \right)^{- 1} \right]^{- 1 / 2}\frac{\partial \mu}{\partial z}   , 
\end{equation}
\begin{equation}
\frac{\partial^2 \mu}{\partial s^2} = \left[ 1 + \frac{z^2}{\alpha^{10 / 3}{R^2} } \left( 1 - \frac{z^2}{\alpha^{4 / 3} R^2} \right)^{- 1} \right]^{- 1 / 2} \frac{\partial}{\partial z} \left(\frac{\partial \mu}{\partial s} \right),
\end{equation}
\begin{equation}
\dfrac{\dd s}{\dd z} = \sqrt{1 + \frac{z^2}{\alpha^{10 / 3} {R^2}} \left( 1 - \frac{z^2}{\alpha^{4 / 3} R^2} \right)^{-1}},
\end{equation}
\begin{equation}
\frac{d r}{d z} = - \frac{z}{\alpha^{5 / 3} R } \left( 1 - \frac{z^2}{\alpha^{4 / 3} R^2} \right)^{- 1 / 2}.
\end{equation}

\subsection{Normal vector to an ellipse surface}
\label{subsec:normal_vec_ellipse}

The general form of the ellipse is 
\begin{equation}
\frac{r^2}{b^2} + \frac{z^2}{a^2} = 1.
\end{equation}
Let
\begin{equation}
F (r, z) = \frac{r^2}{b^2} + \frac{z^2}{a^2}
\end{equation}
be the function that traces the surface of the ellipse. Thus, the normal vector to the surface is $\hat{\bm{n}} = \nabla F/| \nabla F |$. The gradient of $F$ is
\begin{equation}   
\nabla F = \left( \frac{2 r }{b^2}, \frac{2 z}{a^2} \right),
\end{equation}
\begin{equation}
\nabla F = \left( \frac{2 r }{\alpha^{- 2 / 3} R^2}, \frac{2 z}{\alpha^{4 /
   3} R^2} \right),
\end{equation}
and the magnitude of the gradient is
\begin{equation}
| \nabla F | = 2 \alpha^{2 / 3} {R^{- 2}}  (r^2 + z^2 \alpha^{- 4})^{1/2}. 
\end{equation}
Therefore, the $z$-component of the unit vector becomes
\begin{equation}
    n_z = \frac{z}{\alpha^2 (r^2 + z^2 \alpha^{- 4})^{1/2}}.
    \label{sieq:normal_vec_ellipse_temp1}
\end{equation}
Replacing $r$ from \cref{sieq:surf_lapl_temp1} into \cref{sieq:normal_vec_ellipse_temp1} gives
\begin{equation}
    n_z = z \alpha^{- 2} [R^2 \alpha^{-2/3} + z^2 (\alpha^{-4}-\alpha^{-2})]^{-1/2}.
\end{equation}
As expected, if we set $z = \alpha^{2/3} R$, we get $n_z=1$.

\FloatBarrier
\subsection{Parametric analysis}
\label{sisubsec:parametric_analysis}

\begin{figure}[h]
    \centering
    \includegraphics[width=0.5\linewidth]{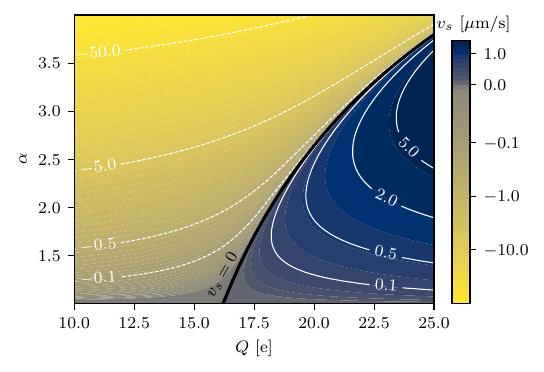}
    \caption{Velocity map of the ellipsoid tip for an $R=2.4$ nm cluster with different $Q$ values. The conditions are $T=150$ K, $\gamma_{\infty}=76$ mN/m, $\delta=-0.8$ \AA, $\rho=1$ g/cm$^3$, $D_s=10^{-15}$ m$^2$/s.}
    \label{fig:vel_map}
\end{figure}

\begin{figure}[!htb]
    \centering
    \includegraphics[width=0.45\linewidth]{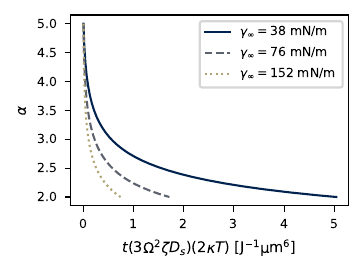}%
    \includegraphics[width=0.45\linewidth]{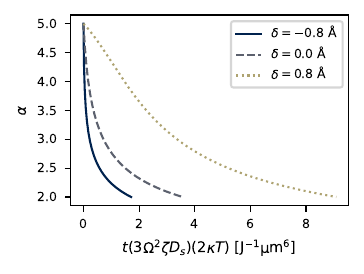}
    \includegraphics[width=0.45\linewidth]{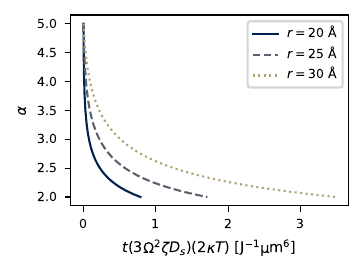}
    \caption{Parametric study of deformation kinetics when an elongated nanocluster collapses to a sphere. The conditions are $\gamma_\infty=76$ mN/m, $\delta=-0.8$ {\AA}, $r=25$ {\AA}, $Q=-10$ e, unless otherwise specified in the legends.}
    \label{sifig:parametric_deformColl}
\end{figure}

\begin{figure}[!htb]
    \centering
    \includegraphics[width=0.45\linewidth]{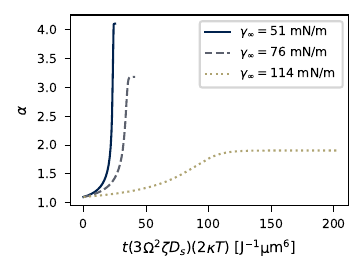}%
    \includegraphics[width=0.45\linewidth]{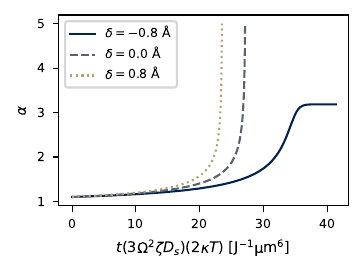}
    \includegraphics[width=0.45\linewidth]{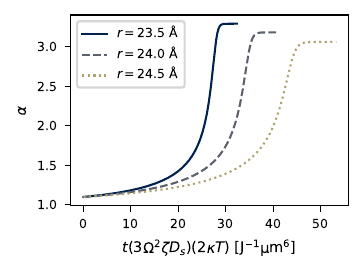}
    \caption{Parametric study of deformation kinetics when a spherical nanocluster elongates. The conditions are $\gamma_\infty=76$ mN/m, $\delta=-0.8$ {\AA}, $r=24$ {\AA}, $Q=-22$ e, unless otherwise specified in the legends.}
    \label{sifig:parametric_deformElong}
\end{figure}

\begin{figure}[!htb]
    \centering
    \includegraphics[width=0.45\linewidth]{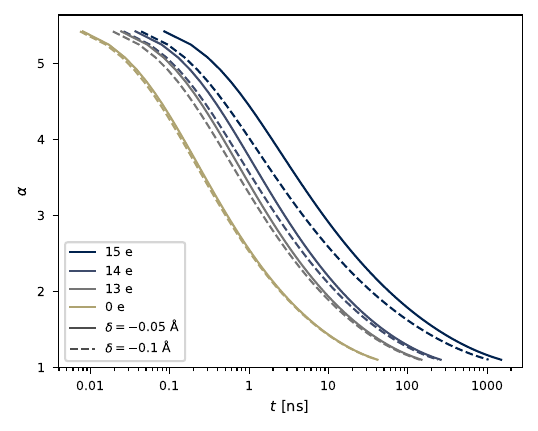}
    \caption{Deformation time of an elongated nanocluster with $R=2.5$ nm, $Q=-16$ e, and initial $\alpha\approx 5.6$, for different $\delta$ values.}
    \label{sifig:md_alpha_time_delta}
\end{figure}

\subsection{Uneven deformation}
\label{sisubsec:uneven_deform}

\begin{figure}[!htb]
    \centering
    \includegraphics[width=0.45\linewidth]{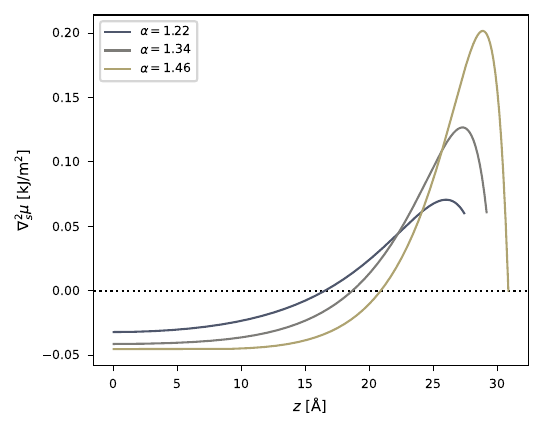}%
    \includegraphics[width=0.45\linewidth]{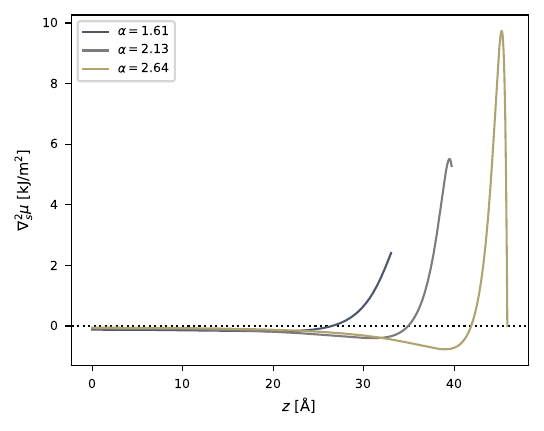}
    \caption{Surface Laplacian of $\mu$ for $Q=-17$ e (left) and $Q=-20$ e (right) at different $\alpha$ values.}
    \label{fig:lbmu_dist_24A}
\end{figure}

\section{Molecular Dynamics simulations}
\label{sisec:atomistic}

\subsection{Simulation details}
\label{sisubsec:md_sim}

The simulation details followed in this study match the ones in Ref. \citenum{Kritikos2026}, and are repeated here. All simulations run on the Large-scale Atomic/Molecular Massively Parallel Simulator (LAMMPS) software \cite{Thompson2022}. Interactions between water molecules are modelled with the SPC/E potential \cite{Berendsen1987}, while electron-water interactions are modelled with the potential detailed in Ref. \citenum{Kritikos2025}. The particle–particle particle-mesh (PPPM) solver is applied to describe long-range electrostatic interactions. The cutoff distance of the PPPM solver is set to 10 {\AA}. The Verlet algorithm is used to propagate the equations of motion with a timestep of 2 fs. 

The preparation procedure is the following: Initially, we create spherical nanoclusters with a radius of 25 {\AA}, containing 2181 water molecules. Then, electrons are placed uniformly near the cluster surface (i.e., on a sphere with a 23 {\AA} radius), using the Fibonacci sphere algorithm \cite{Marques2013}. This was done to approach the charge distribution of spherical conductors. Afterwards, a geometry optimization step was applied to the charged nanoclusters in order to relax the structures. Then, initial velocities are assigned to the atoms and electrons, which were sampled from a Maxwell-Boltzmann distribution at 10 K. The temperature is then ramped up to 220 K for 40 ps. The canonical (NVT) ensemble was used in all simulations. The temperature of the system was regulated using the Nos{\'e}-Hoover thermostat, which rescales the translational degrees of freedom of each particle, and a 200 fs damping constant.

After the preparation of the spherical clusters we follow a strategy of successive cooling-heating cycles to create the elongated clusters. This is done to accelerate the kinetics, and thus observe deformation at MD accessible timescales. At the same time, we ensure that deformation does not happen fast enough so that the cluster forms Taylor cones or breaks apart, instead of forming ellipsoidal structures. The upper temperature limit is set to 220 K, which is above the melting point of 215 K predicted by the SPC/E model \cite{Vega2005}, allowing for fast deformation.

The cluster is frozen based on its deformation kinetics. Specifically, at regular intervals, we compute the radius of gyration of the cluster $r_g^2=\sum_i m_i (\bm{r}_i-\bm{r}_{com})^2/M$, where $m_i$ is the mass of particle $i$, $M$ is the total cluster mass, $\bm{r}$ is the position vector of each particle, and $\bm{r}_{com}$ is the center-of-mass position of the cluster. Then, if $r_g$ increases by more than 15\% or the rate of change of $r_g$ is more than 250 {\AA}/ns, we freeze the cluster to 10 K, we equilibrate the energy for 20 ps, and then ramp up the temperature to 220 K for 100 ps. This process allows us to produce an elongated nanocluster structure, by gradually passing it through successive quasi-equilibrium stages.

The extracted structure is geometry optimized again, and velocities are assigned at 10 K. Then, the cluster is heated to 200 K over 40 ps. For the production simulations, random electrons are selected and removed from the cluster. For each case corresponding to a different number of electrons, 3 different cases were run, differing in the initial velocity distribution and the specific electrons that were removed. Simulations are then performed at a constant temperature of 200 K for 20 ns.

\subsection{Aspect ratio from molecular dynamics simulations}
\label{subsec:clusterdims_MDsims}

To extract the aspect ratio from the MD simulations, we first construct the cluster surface using the alpha-shape method \cite{Stukowski2014} with the OVITO software \cite{Stukowski2010}. According to this method, a spherical probe is used to trace the surface atoms. An iterative scheme then follows to smooth the surface. A probe sphere with a 10 {\AA} radius and a smoothing level of 8 was applied here.

Then, we compute the pair of surface points that are further away from each other, and measure their separation distance. This distance is twice as large as the major radius $a$ of the cluster. We also compute the volume $V$ of the cluster from the constructed surface. Then, we compute $R$ from $R^3=3 V/(4\pi)$ and the minor radius $b$ from $b^2=R^3/a$. Finally, the aspect ratio is computed from $\alpha=a/b$.

\bibliography{references}